\documentstyle[aps,multicol,psfig,eqsecnum]{revtex}  
\input epsf
\def\deltat{\tau}               
\def\dcl{r}
\def\mass{\dcl^{-2}}                    
\def\inte{g}                    
\def\inteIRS{h}                
\def\numcop{{n}}
\def\disfac{\chi}                          
\def\dmhrs{{{\overline{\cal D}}^{\possym}} \Omega }

\def \ofield#1{ \Omega(\hat #1)}
\def\HRS{{\rm HRS}}
\def\WRS{{\rm 1RS}}

\def\action{{\cal F}_{n}^{\rm HRS}}
\def\newaction{{\cal F}_{n}^{\rm 1RS}}
\def\newactionp{{\tilde{\cal F}}_{n}^{\rm 1RS}}
\def\totalaction{{\cal F}_{n}}

\newcommand{\oofield} [2] {\Omega({\hat #1}_{#2})}

\def\possym{\dagger}

\def\sumin{\sum_{j=1}^{N}}

\def\rbv{\hat{e}}

\def\be{\begin{equation}}
\def\ee{\end{equation}}
\def\ba{\begin{eqnarray}}
\def\ea{\end{eqnarray}}
\def\odc{R}
\newcommand{\eqbreak}{
\end{multicols}
\widetext
\noindent
\rule{.48\linewidth}{.1mm}\rule{.1mm}{.1cm}
}
\newcommand{\eqresume}{
\noindent
\rule{.52\linewidth}{.0mm}\rule[-.1cm]{.1mm}{.1cm}\rule{.48\linewidth}{.1mm}
\begin{multicols}{2}
\narrowtext
}
\begin{document} 
\draft 
\title{Density-correlator signatures of the vulcanization transition}
\author{Weiqun Peng and Paul M.~Goldbart} 
\address{Department of Physics, 
University of Illinois at Urbana-Champaign, \\
1110 West Green Street, 
Urbana, Illinois 61801-3080, U.S.A.}
  \date{September 29, 2000}
\maketitle
\begin{abstract} 
Certain density correlators, measurable via various experimental 
techniques, are studied in the context of the vulcanization transition. 
It is shown that these correlators contain essential information
about both the vulcanization transition and the emergent amorphous 
solid state. Contact is made with various physical ingredients that 
have featured in experimental studies of amorphous colloidal and gel
systems and in theoretical studies of the glassy state.
\end{abstract}
\pacs{82.70.Gg, 61.43.-j, 64.70.Pf}	
%
%
%
\noindent
\section{Introduction and basic ingredients}
\label{SEC:Intro}

The vulcanization transition (VT) is an equilibrium phase transition
from a liquid state of matter to an amorphous solid state.  It occurs 
when a sufficient number of permanent random constraints (e.g.~chemical
crosslinks)---the quenched randomness---are introduced to connect the
constituents (e.g.~macromolecules) whose locations are the thermally
fluctuating variables.  A rather detailed description of the VT has 
emerged over the past few years, ranging from a mean-field theory of 
the emerging amorphous solid 
state~\cite{prl_1987,PMGandAZprl,epl,cross,univ} to the critical 
properties of the VT itself~\cite{REF:WPandPG}.  

The purpose of the present Paper is to investigate the properties of
correlators that solely involve the local monomer density, and to
examine the extent to which such correlators provide access to informative
signatures of the
VT and the emergent amorphous solid state.  Along the way, we shall
explore the relationship of these density correlators to various
experimental probes, and also discuss their relationship to the
diagnostics of \lq\lq non-ergodic media\rq\rq\ studied by Pusey, Van
Megen and collaborators in their work on amorphous states of colloidal and gel systems~\cite{REF:PuseyandVanMegen}.  We note that density correlators
closely related to the ones we shall be considering also feature in
certain recent approaches to structural
glasses~\cite{REF:MezardandParisi}, and we shall touch upon the
relationship between our results and those of these recent approaches.

Apart from their connections with related studies by other
researchers, we are motivated to explore the properties of density
correlators in the context of the VT for the following reason.  From
the theoretical perspective, the natural collective coordinate from
which to view the VT is {\it not\/} the local density; rather it is
the amorphous solid order parameter, which becomes nonzero as the
amorphous solid state is entered and whose correlator-decay properties
directly mark the onset of amorphous solidification.  However, as we 
shall discuss further below, from the perspective of experiment, the 
amorphous solid order parameter is rather more elusive than one would 
like, the most direct way to measure it being via incoherent 
quasi-elastic neutron scattering, whereas probes that couple to the
density are more plentiful.  For this reason, we wish to examine
density correlators in the vicinity of the VT, and the extent to which
they can provide access to both the structure of the amorphous solid
state and the long-ranged amorphous solid order-parameter correlations
that develop near the VT.

The approach that we shall adopt to study these density correlators is
based upon a minimal model that takes into account the integrity of
the macromolecules, their thermal position-fluctuations, the
short-range repulsion of the constituent monomers, and the permanent
random constraints imposed by crosslinking.  This minimal model has
previously been shown to give an accurate picture of the universal
properties of the VT, in the sense that its predictions for the
mean-field properties of the amorphous solid state have been verified
in the computer simulations of Barsky and
Plischke~\cite{REF:SJB_MP,REF:WherePub}, and those of its critical
properties that have been elucidated so far (i.e.~the percolative
aspects) are in accordance with the predictions of percolation and
related field-theoretic approaches~\cite{REF:Percolation}.  In order
for our discussion to be concrete and physical, we shall adopt
language specific to randomly crosslinked macromolecular
systems(RCMSs), although our results apply to a broader class of
systems.

Let us now turn to the issue of the order parameter for the VT. 
This order parameter is crafted to detect and diagnose amorphous 
solidification; it is the following function of $\numcop+1$ wavevectors 
$\{{\bf k}^0,{\bf k}^1,\cdots,{\bf k}^{\numcop}\}$:
\begin{equation}
\Big[\,
\frac{1}{N}\sumin 
\int_{0}^{1}ds\,
\big\langle\exp i{\bf k}^{0}\cdot{\bf c}_{j}(s) \big\rangle_{\disfac}
\big\langle\exp i{\bf k}^{1}\cdot{\bf c}_{j}(s) \big\rangle_{\disfac}
\cdots
\big\langle\exp i{\bf k}^{\numcop}\cdot{\bf c}_{j}(s) \big\rangle_{\disfac}
\,\Big], 
\label{EQ:opDefinition}
\end{equation}
where $N$ is the total number of macromolecules, 
${\bf c}_{j}(s)$ (with $j=1,\ldots, N$ and $0\leq s\leq 1$) 
is the position in $d$-dimensional space of the monomer at 
fractional arclength $s$ along the $j^{\rm th}$ macromolecule, $\langle\cdots\rangle_{\chi}$ denotes a thermal average for a 
particular realization $\chi$ of the quenched disorder (i.e.~the 
crosslinking), and 
$\left[\cdots\right]$ represents a suitable averaging over realizations 
of the quenched disorder.   As discussed in detail in Ref.~\cite{cross}, 
this order parameter does indeed detect and diagnose the amorphous 
solid state.  

Why is the amorphous solid order parameter measurable in neutron 
scattering experiments?  In quasi-elastic neutron scattering the 
{\it incoherent} contribution of the scattering cross-section is 
proportional to 
\be
\left\langle
\sum_{j=1}^{N}\int_{0}^{1}ds
\exp\big(i{\bf q}\cdot{\bf c}_{j}(s,0)\big)\,
\exp\big(-i{\bf q}\cdot{\bf c}_{j}(s,t)\big)
\right\rangle_{\disfac}, 
\ee 
where ${\bf c}_{j}(s,t)$ is the position of the monomer at time $t$, 
the $t\to\infty$ limit of the correlator being proportional (up to disorder
averaging) to a special case of Eq.~(\ref{EQ:opDefinition}), viz.,
\be 
\frac{1}{N}\sumin\int_{0}^{1}ds\,
\langle\exp i{\bf q}\cdot{\bf c}_{j}(s)\rangle_{\disfac}\, 
\langle\exp-i{\bf q}\cdot{\bf c}_{j}(s)\rangle_{\disfac}\,\,.
\ee
On the other hand, in several other experimental techniques, 
such as those discussed below, it is some form of correlator involving 
the local monomer density 
\be\rho({\bf r},t)\equiv
\sum_{j=1}^{N}\int_{0}^{1}ds\, 
\delta({\bf r}-{\bf c}_{j}(s,t))\ee
that is probed.  One frequently-measured correlator 
is the auto-correlation function of the local density 
$\langle
\rho({\bf x},0)\, 
\rho({\bf y},t)
\rangle_{\disfac}$, 
or equivalently 
$\langle\rho( {\bf q},0) 
        \rho(-{\bf q},t)\rangle_{\disfac}$, 
where
$\rho({\bf q})$ is the Fourier transform of $\rho({\bf x})$, i.e., 
$\rho({\bf q})=\int d^d x\,\rho({\bf x})
\exp(-i{\bf q}\cdot{\bf x})$~\cite{REF:footnoteaboutq}. For 
example, in neutron scattering experiments this quantity is proportional to 
the {\it coherent} part of the quasi-elastic neutron scattering cross-section 
(see, e.g., Ref.~\cite{cross}, Sec.~IIIE), and in dynamical light scattering 
experiments, such as those performed on \lq non-ergodic\rq\thinspace\ media by 
Pusey and van 
Megen~\cite{REF:PuseyandVanMegen}, this quantity is proportional to the 
intermediate scattering function (also known as the dynamical structure 
factor) $F(k,t)$.  (The average over quenched disorder $[\cdots]$
in the present work essentially plays the role of the ensemble average 
$\langle\cdots\rangle_{\rm E}$ of 
Refs.~\cite{REF:PuseyandVanMegen}.)\thinspace\ 
The present theoretical framework is a static equilibrium framework and, 
as such, is not suitable for computing dynamical correlators.  However, by 
using the cluster property (i.e.~the fact that the connected correlators 
vanish for $t\to\infty$) we see that the long-time limit of the 
density-density auto-correlation function is built from 
the equilibrium entity 
$\langle\rho({\bf x})\rangle_{\disfac}\,
 \langle\rho({\bf y})\rangle_{\disfac}$
or, equivalently, its Fourier transform
$\langle\rho( {\bf q})\rangle_{\disfac}\,
 \langle\rho(-{\bf q})\rangle_{\disfac}$, 
an entity that is calculable (up to disorder averaging) within our 
static equilibrium framework.  In fact, our approach to the VT is 
capable of calculating precisely this kind of quantity and, therefore, 
of providing contact  with experiments.

As our results for density correlators are relatively straightforward,
we first report the results, deferring the construction and operation of the
necessary theoretical machinery to subsequent
sections.  Specifically, we find that:
\hfil\break\noindent
(i)~The usual (i.e.~disorder-averaged) density-density correlator 
$[\langle \rho({\bf q})\rho(-{\bf q})\rangle_{\disfac}]$ is insensitive 
to the VT, depending only analytically on the constraint density, 
both at the level of mean-field theory and beyond (i.e.~to one-loop order).  
\hfil\break\noindent
(ii)~The density-density correlator involving two thermal averages, 
$[\langle\rho( {\bf q})\rangle_{\disfac}\, 
  \langle\rho(-{\bf q})\rangle_{\disfac}]$,   
is zero in the liquid phase but becomes nonzero, continuously, as the 
system enters the amorphous solid phase.  This behavior is a 
manifestation of the freezing-in of random density fluctuations, which 
is the hallmark of the amorphous solid state.  This correlator turns out 
to be proportional the order parameter (at least for weak coupling between the 
density and the order parameter fluctuations). 
As the  order parameter encodes the fraction of localized particles and 
the distribution of localization lengths, this result indicates that these 
physical diagnostics are accessible via this density-density correlator. 
\hfil\break\noindent 
(iii)~The four-density correlator involving two thermal averages, 
$[\langle
\rho( {\bf k})
\rho(-{\bf k})\rangle_{\disfac}\,\langle
\rho( {\bf q})
\rho(-{\bf q})\rangle_{\disfac}]$, 
which can be realized as 
$\lim_{t\to\infty}
[\langle
\rho({\bf k},0)\,
\rho(-{\bf k},0)\,
\rho( {\bf q},t)\,
\rho(-{\bf q},t)\rangle_{\disfac}]$, 
becomes long-ranged as the VT is approached from the liquid side.  
We exhibit this phenomenon at the level of mean-field theory.
\hfil\break\noindent 
(iv)~The two density-chanel signatures of the VT given in (ii) and (iii) also 
provide a means for identifying certain critical exponents at the VT, 
such as the gel-fraction exponent $\beta$, and the correlation-length 
exponent $\nu$.  Therefore, these density signatures provide another 
avenue for accessing experimentally the critical exponents of the VT.

\section{Field-theoretic formulation: 
Minimal model and coupling to density field}

We approach the VT by adopting the spirit of the Landau-Wilson scheme for
continuous phase transitions.  To handle the presence of the random
constraints we invoke the replica trick and adopt the Deam and
Edwards model~\cite{REF:DeamEd} for the statistics of the quenched
randomness (viz.~that the statistics of the random constraints are
determined by the instantaneous correlations of the unconstrained
system).  Thus, we are led to the need to work with the $\numcop \to
0$ limit of systems of $\numcop + 1$ replicas. The additional replica,
labeled by $\alpha = 0$, incorporates the constraint distribution.  With
the effective spatial dimensionality thus being determined to be
$(n+1)d$, symmetry considerations lead to the following minimal
model~\cite{univ}, which takes the form of a cubic field theory
involving an order parameter field $\ofield{k}$ that lives on
$(n+1)$-fold replicated $d$-dimensional space~\cite{REF:Measure}:
\begin{mathletters}
\begin{eqnarray}
f 
&\propto&
-\lim_{n \to 0} n^{-1}{\ln [Z^n]}\,\,,
\\
\left[Z^{n}\right] 
&\propto&
\int \dmhrs 
\exp( - \action), 
\label{EQ:Partition}
\\
\action
\big(\Omega\big)
&=&
N\sum_{\hat{k} \in {\HRS}}
\Big(-a\deltat+\frac{b}{2}|\hat{k}|^2\Big)
\big\vert\ofield{k}\big\vert^{2}
-N\inte
	\!\!\!\!\!\!\!
\sum_{\hat{k}_1,\hat{k}_2,\hat{k}_3\in\HRS}
	\!\!\!\!\!\!\!
\oofield{k}{1}\,
\oofield{k}{2}\,
\oofield{k}{3}\,
\delta_{{\hat{k}_1}+{\hat{k}_2}+{\hat{k}_3},{\hat{0}}}\,.
\label{EQ:LG_longwave}
\end{eqnarray}
\end{mathletters}
Here, $\deltat$ is the VT control parameter, which measures the
reduced density of random constraints, and the coefficients 
$a$, $b$ and $\inte$ depend on the microscopic details of the system. 
We use the symbol 
${\hat k}$ to denote the replicated wavevector 
$\{{\bf k}^0, {\bf k}^1,\ldots, {\bf k}^n\}$, 
and define the extended scalar product
$\hat{k}\cdot\hat{c}$ by 
${\bf k}^0\cdot{\bf c}^0+
 {\bf k}^1\cdot{\bf c}^1+
 \cdots+{\bf k}^n\cdot{\bf c}^n$. 
The symbol $\hat{k}\in{\HRS}$ denotes that the summation over replicated 
wavevectors is restricted to those containing at least two nonzero 
component-vectors ${\bf k}^\alpha$. (We say that this kind of wavevector 
lies in the higher-replica-sector, i.e., the HRS.)\thinspace\ This 
condition on $\hat{k}$ reflects the fact that no crystalline order (or 
any other kind of macroscopic inhomogeneity) is present  or fluctuates critically in the vicinity of the VT.

We now extend the effective free energy, Eq.~(\ref{EQ:LG_longwave}), by 
including the field $\odc$ that is associated with spatial monomer 
density fluctuations.  The field $\odc$ takes as its argument replicated wavevectors 
having exactly one nonzero component-vector ${\bf k}\rbv^\alpha$. 
(We denote by $\{\rbv^{\alpha}\}_{\alpha=0}^{n}$ the collection of unit 
vectors in replicated space, so that, e.g., a generic replicated vector 
$\hat{p}$ can be expressed as 
$\sum_{\alpha=0}^{n}{\bf p}^{\alpha}\rbv^{\alpha}$.)
We term the subset of replicated wavevectors having exactly one nonzero 
component-vector the one-replica-sector (\WRS) of wavevectors; 
we term the corresponding fields $\odc({\bf k}\rbv^\alpha)$ \WRS\ fields.
We extend the effective Landau free energy, Eq.~(\ref{EQ:LG_longwave}), 
by incorporating the \WRS\ fields~\cite{REF:WRSMeasure}, which represent local 
density fluctuations, and add the significant symmetry-allowed cubic term 
that couples the order-parameter and density fields, thus arriving at 
\begin{mathletters}
\begin{eqnarray}
\totalaction \big(\Omega, \odc\big)
&=&\action \big(\Omega\big) + \newaction\big(\odc\big)
-{\inteIRS \over{N}} 
	\!\!
\sum_{\hat{k}_3\in\HRS \atop {\hat{k}_1,\hat{k}_2\in\WRS}}
	\!\!\!
\odc({\hat k}_{1})\,
\odc({\hat k}_{2})\,
\Omega({\hat k}_{3})\,
\delta_{{\hat{k}_1}+{\hat{k}_2}+{\hat{k}_3},{\hat{0}}}\,,
\label{EQ:LG_WRS_HRS1}
\\
\newaction\big(\odc\big)
&=&
{1\over{N}}\sum_{\hat{k} \in {\WRS}}
\Big(\mass+\frac{c}{2}|\hat{k}|^2\Big)
\big\vert \odc(\hat k)\big\vert^{2} +\cdots\,\,\,.
\label{EQ:LG_WRS_HRS2}
\end{eqnarray}
\end{mathletters}
The term $\newaction\big(\odc\big)$ is the effective free energy for 
the density fluctuations; in principle, it also includes non-linear 
couplings between the $\odc$ fields.  This effective free energy term 
already incorporates the effects of the short-range repulsion between macromolecules.
The parameter $\dcl$ is the correlation length for density fluctuations.  
(In the context of a dense melt, it is simply determined by the monomer 
density and the effective excluded-volume interaction 
strength~\cite{REF:DoiEdwards}.)\thinspace\ The correlation length $\dcl$ remains large and varies analytically (with the 
constraint density) across the VT, and the $\odc$ field remains a non-ordered 
field.  This is representative 
of the fact that the {\it disorder-averaged} physical monomer-density is 
homogeneous in both the liquid state and the amorphous solid state. 

In addition to the coupling presented in Eq.~(\ref{EQ:LG_WRS_HRS1}),
there is one further term at cubic order that couples the $\odc$
and $\Omega$ fields, i.e., the vertex $\Omega\Omega\odc$ consisting of two $\HRS$ fields and one $\WRS$ field.  It can readily be shown by 
dimensional analysis that both this cubic vertex and that given in 
Eq.(~\ref{EQ:LG_WRS_HRS1}) are
irrelevant with respect to the fixed points of \HRS\ $\Omega$ theory
near $d=6$ and, therefore, the critical properties of the VT (i.e.~the 
fixed point structure, the flow equation and the critical exponents) 
are not affected by the coupling to density fluctuations(at least near 
$d=6$).  Based on the effective free energy~(\ref{EQ:LG_WRS_HRS1}), 
our approach is to explore the correlators of the \WRS\ fields 
(and hence the density correlators), taking into account the effects 
of the VT in the \HRS\ fields by treating what happens in the
\HRS\ as \lq\lq input\rq\rq\ to be added to the effective free 
energy of the \WRS\ theory, and working perturbatively 
(i.e.~effectively we assume that $\inteIRS$ is small).

The reason that we ignore the the cubic coupling $\Omega\Omega\odc$,
besides its irrelevance in the renormalization-group sense, is that 
it does not contribute to the density correlators that we are 
interested in (at least to one-loop order).  There are two points to 
make in this regard.
First, at the mean-field level, the
$\HRS$ field can be viewed as an external source for the $\WRS$ field
in the cubic coupling $\Omega\Omega\odc$. Due to translational
invariance, $\langle\oofield{k}{1}\oofield{k}{2}\rangle^{\action}=0$.
(We use $\langle\cdots\rangle^{\action}$ to denote a statistical
average weighted by ${\action}$.)\thinspace\ Therefore, on average,
the term $\Omega\Omega\odc$ will not generate a non-zero
$\langle\odc\rangle$ [$\langle\cdots\rangle$
denotes an average weighted by the replicated effective free energy
presented in Eq.~(\ref{EQ:LG_WRS_HRS1})]. 
Second, at the one-loop level
(and beyond), this term will renormalize the coefficient $\mass$ (in a
singular way) but, as has already been shown in App.~B of
Ref.~\cite{REF:WPandPG}, at least to the one-loop level, there is (in
the replica limit) no contribution to the density-density correlator
coming from \HRS\ critical fluctuations via this kind of vertex.

In order to help make the physical content of the results that we 
shall present clear, we pause to give the relationship between the
physical density correlators and the $\odc$ correlators: 
\begin{eqnarray} 
\lim_{n\to 0}\left\langle\odc( {\bf k}\rbv^{\alpha})\, 
\odc(-{\bf k}\rbv^{\beta})\right\rangle_{\rm c}
=\cases{
\left[\left\langle\rho( {\bf k})\,
                  \rho(-{\bf k})\right\rangle_{\disfac}\right];
&for $\alpha=\beta$,\cr
\noalign{\smallskip}
\left[\,
\left\langle\rho( {\bf k})\right\rangle_{\disfac}\, 
\left\langle\rho(-{\bf k})\right\rangle_{\disfac}
\,\right];
&for $\alpha\ne\beta$,
\label{EQ:Offdiagcorr}
\cr}
\end{eqnarray}
where ${\rm c}$ denotes that a 
correlator is connected. (Such connections can be  
established by following the replica technique that
is used in Appendix A of Ref.~\cite{cross}.)  
On the right hand side of Eq.(~\ref{EQ:Offdiagcorr}), 
the correlators differ in the locations 
of the thermal averages; on the left hand side they differ in 
their replica indices, the former being diagonal and the latter 
being off-diagonal in replica space.

\section{Freezing-in of density fluctuations}

Now that we have constructed an extended model containing not only
the critical order parameter (i.e.~\HRS) fields but also the
noncritical replicated density (i.e.~\WRS) fields, we proceed to
study the effect of critical \HRS\ phenomena on the density fields,
treating the latter at the tree level.  The basic mechanism at work is
that the order parameter field, which is capable of ordering
spontaneously, couples to the density fluctuations via a cubic vertex
that is replica-off-diagonal as far as the density fields are
concerned.  A non-zero value of
${\overline\Omega}\equiv\langle\Omega\rangle^{\action}$, as occurs in
the amorphous solid state due to spontaneous symmetry breaking in the
\HRS, contributes replica-off-diagonal terms to the \lq\lq mass matrix\rq\rq\ 
of the $\odc$-field and, hence, leads to the
existence of nonzero replica-off-diagonal density-field correlators.

In order to see this more clearly, we replace $\Omega$ by its
expectation value plus fluctuations, i.e., we write 
$\Omega={\overline \Omega}+{\delta\Omega}$ 
and, hence, arrive at the effective free energy
\begin{mathletters}
\begin{eqnarray}
\totalaction \big(\odc,\Omega\big)
&=&\action \big(\Omega\big)+\newactionp \big(\odc,{\overline \Omega}\big)
-{\inteIRS\over{N}} 
\sum_{\hat{k}_3\in\HRS \atop {\hat{k}_1,\hat{k}_2\in\WRS}}
\odc({\hat k}_{1})\,
\odc({\hat k}_{2})\,
{\delta\Omega}({\hat k}_{3})\,
\delta_{{\hat{k}_1}+{\hat{k}_2}+{\hat{k}_3},{\hat{0}}}\,,
\\
\newactionp\big(\odc,{\overline \Omega} \big)
&=&
{1\over{N}}
\sum_{{\bf k}\neq {\bf 0}}
\sum_{{\alpha, \beta =0} \atop {(\alpha \neq \beta})}^{n}
\left( (\mass+
  {\frac{c}{2}}k^2) 
  \delta^{\alpha, \beta}  
-\inteIRS\,{\overline \Omega} (-{\bf k}\rbv^1+{\bf k}\rbv^2)
(1-\delta^{\alpha, \beta})\right)
\odc({\bf k} \rbv^{\alpha})
\odc(-{\bf k} \rbv^{\beta}) +\cdots\,.
\label{EQ:hasmassmatrix}
\end{eqnarray}
\end{mathletters}
To arrive at this result, we have taken advantage of the facts that 
both $\totalaction$ and ${\overline\Omega}$ are replica (i.e.~permutation) 
symmetric, and that
${\overline\Omega}(-{\bf k}_1 \rbv^{\alpha}-{\bf k}_2 \rbv^{\beta})$  
is macroscopically translational invariant (i.e.~it contains a 
factor of $\delta_{{\bf k}_1 +{\bf k}_2,{\bf 0}}$)~\cite{univ}. 

We now aim to compute the correlator 
$\left\langle\odc( {\bf k}\rbv^{\alpha})\, 
             \odc(-{\bf k}\rbv^{\beta}) \right\rangle$.
By treating $h$ as a small quantity and expanding perturbatively, 
a direct calculation yields 
\be
\left\langle
        \odc( {\bf k}\rbv^{\alpha})\, 
        \odc(-{\bf k}\rbv^{\beta})\right\rangle =
\left\langle
        \odc( {\bf k}\rbv^{\alpha})\, 
        \odc(-{\bf k}\rbv^{\beta})\right\rangle^{\newactionp} 
        +{\it O}(h^2). \ee 
To obtain the correlator at the tree  (in $\odc$) level, we neglect the
nonlinear self-couplings of $\odc$ and then invert the coefficient 
matrix of the quadratic term in Eq.~(\ref{EQ:hasmassmatrix}).  Thus, in the $n\to 0$ limit, 
we arrive at
\be
\lim_{n\to 0}
\left\langle 
\odc( {\bf k}\rbv^{\alpha})\,
\odc(-{\bf k}\rbv^{\beta})
\right\rangle 
=\cases{ {\displaystyle
\frac{N}{2\mass+ck^2}}, 
&for $\alpha=\beta$;\cr 
{\displaystyle \frac {1}{2}\frac{N \inteIRS 
{\overline \Omega} (-{\bf k}\rbv^1+{\bf k}\rbv^2)}
{(\mass+ \frac{1}{2}c k^2)((\mass+ \frac{1}{2}c k^2)
+\inteIRS {\overline \Omega} (-{\bf k}\rbv^1+{\bf k}\rbv^2))}}, 
&for $\alpha\ne\beta$.\cr}
\label{EQ:corrresults}
\ee
In the liquid state we have ${\overline\Omega}=0$, and therefore 
$\langle\odc( {\bf k}\rbv^{\alpha})\, 
        \odc(-{\bf k} \rbv^{\beta})\rangle =0$ (for $\alpha\ne\beta$). 
However, in the amorphous solid state 
${\overline\Omega}({\bf k}_1 \rbv^{1}+{\bf k}_2 \rbv^{2})=
q\delta_{{\bf k}_1+{\bf k}_2,{\bf 0}}\, 
w( k_1^2 + k_2^2)\ne 0$, and therefore  
$\langle\odc( {\bf k}\rbv^{\alpha})\, 
        \odc(-{\bf k} \rbv^{\beta})\rangle \ne0$ (for $\alpha\ne\beta$). 
Here, the number $q$ is the gel fraction and the function $w(|{\hat k}|^2)$, 
which decays rapidly with increasing wavevector magnitude on the 
characteristic wavevector scale $\deltat^{\nu}$~\cite{REF:WPandPG}, 
encodes the distribution of localization lengths.  The simplest 
setting for the density-density correlator emerges near the 
VT, where $q$ is small and $w(|{\hat k}|^2)$ is negligible 
unless $|{\hat k}|\alt\deltat^{\nu}$.  In this regime, by making 
use of Eq.~(\ref{EQ:Offdiagcorr}) we find that
\begin{equation} 
\left[\,
\left\langle\rho( {\bf k}) \right\rangle_{\disfac}\,
\left\langle\rho(-{\bf k}) \right\rangle_{\disfac}
\,\right]
=\cases{%
        {\displaystyle 0}\smallskip,       &liquid state;\cr
        {\displaystyle (N\inteIRS\dcl^{4}/2)
        {\overline\Omega}
        (-{\bf k}\rbv^{1}+{\bf k}\rbv^{2}) 
=       (N\inteIRS\dcl^{4}/2){\deltat}^{\beta}\, 
        w(k^{2}\deltat^{-2\nu})}, &amorphous solid state.}
\end{equation}
On the other hand, the diagonal correlator 
$\big[
\left\langle\rho( {\bf k}) 
            \rho(-{\bf k}) 
\right\rangle_{\disfac}
\big]$
does not vary with $\deltat$ (and hence varies smoothly with the 
physical constraint density). 

Deeper into the amorphous solid state, the order parameter 
${\overline\Omega}$ does not decay so rapidly with $k$ and hence 
the quantity 
$\big[\,
\left\langle\rho( {\bf k})\right\rangle_{\disfac}\, 
\left\langle\rho(-{\bf k})\right\rangle_{\disfac}
\,\big]$
should remain appreciable (and thus experimentally accessible) over a
wider range of $k$.  Now, we expect Eq.~(\ref{EQ:corrresults}) to
remain valid, provided the coupling $h$ is small and both the 
wavevector dependence of $h$ and the finer wavevector dependence 
of \WRS\ bare correlator are incorporated. (We have omitted the 
wavevector dependence of $h$ so as 
to simplify our presentation.)\thinspace\  Under these circumstances, 
the wavevector dependence of the replica-off-diagonal correlator has 
the possibility of exhibiting additional features, representative of 
the ordinary density-density correlator  
$\big[
\big\langle\rho( {\bf k}) 
            \rho(-{\bf k}) 
\big\rangle_{\disfac}
\big]$, superposed on the decaying trend due to the factor $w$ (i.e.~due to random monomer localization)~\cite{REF:PVMnote}.

\section{Inherited criticality of density correlators}

In the \HRS\ field theory, the VT is signaled in two ways: 
(i)~via the emergence of a nonzero order parameter, and 
(ii)~via the divergence of the correlation length of order-parameter fluctuations.  We have already studied 
the replica-off-diagonal density correlator, which is closely 
related to the order parameter and becomes nonzero upon entering the 
amorphous solid state.  We now examine the four-field density 
correlator
$[\langle
\rho( {\bf k})\,
\rho(-{\bf k})\rangle_{\disfac}\,\langle
\rho( {\bf p})\,
\rho(-{\bf p})\rangle_{\disfac}],$
which has the property that it becomes long-ranged at the VT. 

\begin{figure}[hbt]
\epsfxsize=3.5in
  \centerline{\epsfbox{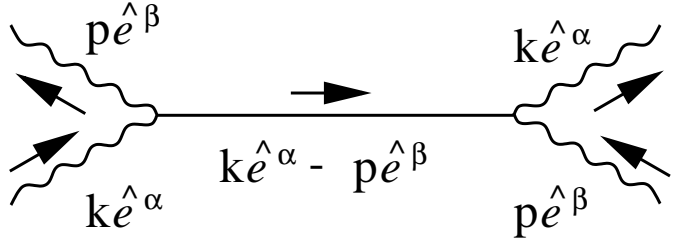}} 
\vskip0.50cm
\caption{Divergent 4-field tree level density correlator.
Solid line indicates the bare \HRS\ correlator; 
wavy lines indicate bare \WRS\ correlators.
\label{FIG:4_density}}
\end{figure}
We calculate the correlator mentioned in the previous sentence 
at the tree level (with respect to $h$ 
vertices) in the liquid state and at the VT itself: the Feynman 
diagram shown on Fig.~\ref{FIG:4_density} is the only contribution, and
gives
\begin{eqnarray} 
&&
[\langle
\rho( {\bf k})\,
\rho(-{\bf k})\rangle_{\disfac}\,\langle
\rho( {\bf p})\,
\rho(-{\bf p})\rangle_{\disfac}]=
\lim_{n\to0}
\langle
\odc( {\bf k} \rbv^{\alpha}) \odc(-{\bf p}\rbv^{\beta})
\odc(-{\bf k} \rbv^{\alpha}) \odc({\bf p}\rbv^{\beta})
\rangle
\nonumber
\\
&&\qquad\qquad
\propto (h/N)^2
\Big\{
\langle\odc({\bf k}\rbv^\alpha)
       \odc(-{\bf k}\rbv^\alpha)\rangle^{\newaction}
\Big\}^2
\Big\{
\langle
\odc({\bf p}\rbv^\beta)
\odc(-{\bf p}\rbv^\beta)
\rangle^{\newaction}\Big\}^2
\langle\Omega({\bf k}\rbv^\alpha-{\bf p}\rbv^\beta)
       \Omega(-{\bf k}\rbv^\alpha+{\bf p}\rbv^\beta)
       \rangle^{\action}.
\end{eqnarray}
As anticipated, this density correlator becomes long-ranged at the VT, 
due to the factor of the \HRS\ order-parameter correlator 
$\langle\Omega\Omega\rangle^{\action}$ which, itself, becomes 
long-ranged at the VT.  

\section{Discussion and conclusions}

We have studied density-sector correlators that furnish analogs of
the two principal order-parameter signatures of the VT: the
off-diagonal density correlator 
$[\langle\rho( {\bf q})\rangle_{\disfac}\,
\langle\rho(-{\bf q})\rangle_{\disfac}]$, 
which becomes nonzero as the amorphous solid state is entered; 
and the four-field density correlator
$[\langle
\rho( {\bf k})\,
\rho(-{\bf k})\rangle_{\disfac}\,\langle
\rho( {\bf p})\,
\rho(-{\bf p})\rangle_{\disfac}]$, 
which becomes long-ranged at the VT.  We have shown that these density
correlators provide useful information about both the emergent
amorphous solid state and the critical properties of the transition
itself.  They provide schemes for accessing experimentally the 
kinds of quantities that have been found useful in theoretical 
investigations of the liquid, critical and amorphous solid
states, e.g., of vulcanized matter.  We are not aware of any 
explorations of such signatures in the density correlators via  percolation/gelation approaches to the VT.

Although, as we have seen, the off-diagonal density correlator 
$[\langle\rho( {\bf q})\rangle_{\disfac}\, 
  \langle\rho(-{\bf q})\rangle_{\disfac}]$ 
is closely related to the order parameter of the VT, its non-zero value
being induced by a non-zero order parameter, the off-diagonal density
correlator can be used to diagnose a more general class of systems.
By looking at the microscopic definition of the order parameter for the 
VT, Eq.~(\ref{EQ:opDefinition}) (especially for the special case in which 
only two thermal averages are involved), we see that it involves only a 
{\it single\/} summation over monomers and, therefore, the order parameter
cannot be expressed in terms of local monomer densities. In essence, the
order parameter signals the amorphous solid phase via its detection of 
all individual monomers that are localized (and, owing to the explicit 
crosslinks, distinguishable).  On the other hand, the off-diagonal 
density correlator 
$[\langle\rho( {\bf q})\rangle_{\disfac}\, 
  \langle\rho(-{\bf q})\rangle_{\disfac}]$ 
signals the amorphous solid state via the detection of the frozen 
structure of the local density fluctuations.  
In RCMSs, the localization of monomers invariably induces the freezing
of this structure, so these two quantities are equivalent.  However in 
systems such as glasses, the (local density) structure is, presumably,
frozen but each individual particle is able to wander throughout the 
system, given enough time, and therefore all particles retain their  indistinguishability~\cite{REF:Mezard}.  A quantity essentially
identical to the off-diagonal correlator considered here has 
been employed by Mezard and Parisi~\cite{REF:MezardandParisi} in the 
context of their theoretical approach to glassy systems.  A
similar quantity is employed in the diagnosis of the freezing-in of 
structure in colloidal glassy systems and gel systems in the dynamical light scattering experiments of Pusey and Van Megen and 
collaborators~\cite{REF:PuseyandVanMegen,REF:PuseyandVanMegen2}

Vulcanized matter certainly differs from glassy systems, inasmuch as
it possesses explicit quenched disorder in the form of permanent
random crosslinks.  Because of this, it affords a framework in which
the off-diagonal density correlator and other density correlators are
directly calculable in a well controlled way, both near the transition
and in the amorphous solid state. Despite the difference with glassy systems, we hope
that results such as those presented here in the setting of
vulcanized matter will not only be useful but also shed 
some light on the more difficult
problem of glassy systems.

\noindent
\section*{Acknowledgments}
\label{SEC:Acknowledgments}
We thank Avi Halperin, Vincent Liu, Marc M\'ezard, Jos\'e Mar\'{\i}a Rom\'an and,  
especially, Eduardo Fradkin for useful discussions.  
This work was supported by the U.S.~National Science 
Foundation through grant DMR99-75187. 
  

\begin{references}
\bibitem{REF:Measure}
The measure $\dmhrs$ indicates that we integrate $\ofield{k}$ with
$\hat k \in {\HRS}$ and with $\hat k$ restricted to a half-space via
an additional condition $\hat k \cdot \hat m > 0$ for some suitable 
unit $(n+1)d$-vector $\hat m$.  This is due to the fact that 
$Q(-\hat k)=Q(\hat k)^{\ast}$ (i.e.~its Fourier transform is real).
\bibitem{REF:WRSMeasure}
Functional integration over \WRS\ fields is implied.
\bibitem{REF:PVMnote}
Such additional features have been observed in amorphous 
colloidal systems; see Ref.~\cite{REF:PuseyandVanMegen2}.
\end{references}
\end{document}